\documentclass{cernrep}

\usepackage{amsmath,amssymb,amsfonts,color,graphicx,cite}
\usepackage{placeins}

\newcommand{\be}{\begin{equation}}
\newcommand{\ee}{\end{equation}}
\newcommand{\bea}{\begin{eqnarray}}
\newcommand{\eea}{\end{eqnarray}}

\newcommand{\tev}{\,\, \mathrm{TeV}}
\newcommand{\gev}{\,\, \mathrm{GeV}}

\renewenvironment{thebibliography}[1]{%
  \begin{oldthebibliography}{#1}%
    \setlength{\parskip}{0ex}%
    \setlength{\itemsep}{0ex}%
}%
		 {%
  \end{oldthebibliography}%
		 }

\title{Higgs boson discovery versus sparticles prediction:\\
Impact on the pMSSM's posterior samples from a Bayesian global fit.} 
\author{
Shehu~S.~AbdusSalam$^1$, 
Debajyoti~Choudhury$^2$
}
\institute{
$^1$ The Abdus Salam International Centre for Theoretical Physics,
  Strada Costiera 11, Trieste 34014, Italy\\ 
$^2$ Department of Physics and Astrophysics, University of Delhi, 
  Delhi 110007, India
}

\begin{document}

\maketitle 

\noindent

\begin{abstract}
The signal strength of the recently discovered Higgs boson-like
particle in the diphoton channel seemingly constrains physics beyond
the standard model to a severe degree.  However, the reported signal
strength is prone to possible underestimation of uncertainties. We
propose a discriminant that is relatively free of many of the
theoretical uncertainties, and use this to gauge the impact on
  the phenomenological MSSM.  A Bayesian global fit to all the pre-LHC
  data results in posterior distributions for the masses that are
  neither very restrictive, nor sufficiently prior-independent (except
  for the Higgs and stop masses). The imposition of the Higgs data, on
  the other hand, yields interesting and nearly prior-independent
  constraints.  In particular, the existence of some light
  superpartners is favoured.\\
  \vspace{1.0cm}
\end{abstract}

\section{Introduction}
\label{sec:intro}

The recent discovery of a 126.5 GeV or 125.0 GeV Higgs boson-like 
state, respectively by the ATLAS and CMS 
collaborations~\cite{:2012gk,:2012gu}, at the LHC marks the beginning
of an exciting scientific era for understanding the mechanism
responsible for electroweak symmetry breaking and for endeavour in
search for new, beyond the standard model (BSM), physics.
For a standard model-like Higgs boson, the search strategies at the LHC
involve, in order of decreasing sensitivity, production processes 
driven by gluon fusion $gg \rightarrow H$, vector boson
fusion $q\bar{q} \rightarrow Hqq$, associated production with a vector
boson $q\bar{q} \rightarrow V H$ and the associated production with a
top-quark pair $gg \rightarrow t\bar{t}H$.
Further to production, the main search channels are
defined by decay modes into $\gamma \gamma$, $ZZ^{*}$, $WW^{*}$,
$b\bar{b}$ and $\tau^+ \tau^-$. Subsequent decays of the $W$-
and $Z$-bosons with at least one leptonic decay are also considered.
For the experiments with the Tevatron collider, the
main production modes, for the most sensitive search topologies, are
the associated production of a Higgs with a 
vector boson $q\bar{q} \rightarrow V H$, with the subsequent decay of the
Higgs into a $b\bar{b}$ pair, and the $gg$-fusion production with the 
subsequent decay into two W bosons which then decay leptonically.
The LHC discovery has supporting evidence from the Tevatron
experiments (see e.g.~\cite{Oksuzian:2012rz} and references therein)
where the excess seen is in accord with a 115 to 135 GeV 
Higgs boson.

Usually, the Higgs boson signal is quantified by the 
production cross section times the relevant decay-channel branching 
fraction to give the {\em signal strength}, 
\be   \label{muX}
\mu_X = \frac{ \sigma(gg \rightarrow h) \, Br(h
  \rightarrow X)}{ \sigma(gg \rightarrow h)_{SM} \, Br(h
  \rightarrow X)_{SM}}.
\ee 
Here $X = \gamma \gamma, ZZ^{*}, WW^{*}, b\bar{b}$, or $\tau^+
\tau^-$. Of the Higgs boson decay channels reported by the
experimental collaborations, $\mu_{\gamma \gamma}$ apparently makes
for a very strong BSM-constraining observable.  However, as argued in
Ref.~\cite{Baglio:2012et}, it is relevant to question whether the
apparent excess is actually due to new physics effect or due to
under-estimations in theoretical uncertainties. The question would
also propagate directly to any analysis that employs the reported
signal strength. In order to avoid this and for the purpose of
allowing robust conclusions to be derived from the discovery data, we
are going to propose a discriminant between new physics and the
standard model (SM), which should be largely free of the mentioned
source of uncertainties.  

We consider, here, the R-parity conserving minimal supersymmetric
standard model (MSSM) as a template for BSM. Rather than appeal to a
particular scenario for supersymmetry breaking and/or make an Ansatz
for the parameters at some high scale, we, instead, use the
phenomenological MSSM (pMSSM) parametrisation
\cite{Haber1998,Djouadi:1998di} to characterise the supersymmetric
sector.  Although this is almost the most general scenario given the
filed content, it is, nonetheless, subject to very many
constraints. For example,
Refs.\cite{AbdusSalam:2008uv,AbdusSalam:2009qd} had used all pre-LHC
data to arrive at Bayesian posterior distributions for the
superpartner masses. We discuss here how the mass of the Higgs
resonance and the information on the signal strengths (in particular,
the robust discriminant) can be used to further refine the posterior
to yield some very interesting clues about the pMSSM spectrum.  Other
discussions concerning LHC constraints on the pMSSM were effected in
Refs.~\cite{Conley:2011nn,Allanach:2011ej,Sekmen:2011cz,Arbey:2011un,AlbornozVasquez:2011aa}.

Several studies have been made of 
the  MSSM Higgs
sector in light of the observed excess (Refs.~\cite{Djouadi:1996pb,Djouadi:1998az,AlbornozVasquez:2011aa,Arbey:2012dq,Carena:2012gp,Cao:2012fz,Cao:2012yn}
comprise a non exhaustive list).
 in the diphoton channel signal
strength or the effect of light sparticles in altering it.
Some of these
attempts are based on
random scans in  the pMSSM parameter space
while others 
restrict themselves to  sub-spaces or hyper-planes
of the MSSM parameters space. In this article we make use of a complete set of
pMSSM points that came out of statistically convergent Bayesian
fits. The aim is to derive the implications of the recent LHC
discovery on the pMSSM posterior distributions by gauging its
effect on the sparticle masses.
In the
the rest of this section we introduce the BSM
discriminant for the analysis. In Section~\ref{sec:pmssm} we give a
brief recapitulation of the pMSSM's posterior distributions for
sparticles masses, Higgs boson signal strengths, and the proposed
discriminant predictions from a pre-LHC Bayesian fit to
data. Section~\ref{sec:impact} describes the implementations and
implications of the Higgs boson data on the pMSSM predictions.

\subsection{A useful discriminant}
\label{sec:discrim}
The ratios $\mu_{X}$, see eq.(\ref{muX}), represent good measures for
possible deviations from the SM and have, indeed, been exploited to
discriminate against BSM models. However, such an exercise ought to be
approached with care, for it is yet early days of Higgs physics and
the statistics are still low. A naive adoption of the current bounds
would indicate a significant tension with the SM
itself~\cite{Eberhardt:2012gv}. Of particular concern are the
theoretical errors pertaining to the choice of the parton
distributions and the renormalization ($Q_R$) and factorization
($Q_F$) scales. These uncertainties can be large and have been
addressed in Ref.~\cite{Botje:2011sn}. The LHC Higgs cross section
working group recommends~\cite{Dittmaier:2011ti} that in convoluting
the errors, these should be added linearly. However, both ATLAS and
CMS have added them in quadrature. As Ref.\cite{Baglio:2012et}
explicitly demonstrates, adopting the more conservative approach of
linear addition would substantially increase the error bars to the
extent of a $1\sigma$ agreement with the SM. A further uncertainty
pertains to the use of an effective field theory (in the limit of an
infinitely heavy top) used for many of the higher-order calculations.

This criticism brings into question the naive use of $\mu_{X}$. However,
note that the uncertainty applies primarily to the production process
under consideration viz. $p p \to h$ (driven, at the lowest order,
through gluon-fusion). In fact, as far as non-hadronic final states
(diphoton, and four-lepton, charged or neutral) are concerned, this is
an exact statement as quantum chromodynamics (QCD) corrections can occur exclusively in the
initial state alone. For $p p \to h \to b \bar b$, though, one could
have gluon exchanges between the initial and final state particles,
thereby engendering non-universal dependence on the parton density.
Nevertheless, as is well-known, such diagrams result in only a 
subdominant component of the higher-order QCD corrections, and, thus,
even for the $p p \to h \to b \bar b$ case, the dependence is 
identical to a very good approximation. The same argument holds for 
$p p \to h \to V V^*$ with the gauge boson(s) going into jets.

In other words, if we consider only those final states that are primarily
driven by Higgs-production through gluon-gluon fusion,
the ratios of ratios\footnote{Ref.\cite{Arvanitaki:2011ck} too had 
raised the possibility of using such ratios, without putting it into
practice. Refs~\cite{Guasch:2001wv,Assamagan:2004wq,Assamagan:2004ji}
also used similar ratios for suppressing systematic errors.}, namely, 
\begin{equation}
   R_{ij} \equiv \frac{\mu_i}{\mu_j} \ ,
\end{equation}
are essentially free of the
aforementioned uncertainties due to the choice of $Q_{R,F}$ as well as
the choice of parton distributions\footnote{It
  is also worthwhile to note that even the ratios of, say, gluon-fusion
  and vector-boson fusion cross-sections also profit on this account,
  though not to the same high degree as is applicable 
  here.}. Here $i$ and $j$ represent distinct choice from $X =
\gamma \gamma, ZZ^{*}, WW^{*}, b\bar{b}$, or $\tau^+ \tau^-$. 
The same argument would hold for other systematic errors
that are factorisable, or are nearly universal. A prime example 
is afforded by the uncertainty in the luminosity measurement. 

Thus, it makes eminent sense to consider all the independent ratios
$R_{ij}$, and, at best, one individual $\mu_i$, for, together, they
embody the same information as the set $\{\mu_i\}$. There is, of
course, a price to pay: the relative statistical errors in $R_{ij}$
would be larger than those in the individual $\mu_i$. Furthermore,
certain statistical errors in $\mu_i$ are correlated\footnote{An
  example would be those pertaining to the electromagnetic calorimeter
  measurements, relevant  
to both the diphoton channel and the $VV^*$ modes containing electrons.}, 
and, unless the
correlation matrix is available, a naive combination would tend to
overestimate the error in $R_{ij}$.  This, certainly, is the case in
the current context in the absence of sufficient public information 
about the ATLAS and CMS analyses. 

Nonetheless, for our analyses we shall consider such ratios
$R_{ij}$. Of particular interest  
here is the ratio of the two reported discovery channels, namely 
four-lepton and diphoton. The ATLAS and CMS ratios for the same, as
shown in Table~\ref{muus}, are consistent 
with each other within about $0.8 \sigma$, and combining the two, we have,
\begin{equation} 
    R_{ZZ \, ; \gamma\gamma} \equiv \frac{\mu_{ZZ}}{\mu_{\gamma\gamma}}
                            = 0.56 \pm 0.25 \ , 
    \label{R_zz_diphot}
\end{equation}
which, in itself, is
consistent with the SM within $2 \sigma$. Yet, the very fact
that the central value of $R_{ZZ \, ; \gamma\gamma}$ is substantially
below the SM expectation would turn out to have rather interesting
consequences.  It should be realized that this deviation is primarily
driven by the fact that $\mu_{\gamma \gamma}$ is significantly larger
than unity; yet, unlike the latter, this observable is largely free
from the aforementioned theoretical errors. Moreover, the error in 
eq.(\ref{R_zz_diphot}) has been somewhat overestimated and would be 
reduced substantially once the error correlation matrix is available.
This renders any inference drawn from the use of eq.(\ref{R_zz_diphot})
to be conservative in nature. 

\begin{table}
\begin{center}
\begin{tabular}{|c|l|l|}
\hline \hline
Search channel & ATLAS $\mu$ at $m_h = 126.5 \gev$ & CMS $\mu$ at $m_h
= 125.0 \gev$ \\ 
\hline
 $H \rightarrow \gamma \gamma$ & $1.8 \pm 0.5$, $4.5 \sigma$  & $1.6
\pm 0.4$, $4.6 \sigma$ \\
 $H \rightarrow ZZ^{*} $ & $1.2 \pm 0.6$, $3.6 \sigma$ &
$0.7^{+0.4}_{-0.3}$, $3.2 \sigma$ \\ 
\hline\hline
\end{tabular}
\caption{\normalsize The ATLAS~\cite{:2012gk} and CMS~\cite{:2012gu} magnitudes
  for the Higgs boson production signal strengths.} 
\label{muus}
\end{center}
\end{table}

Similarly, we also have
\begin{equation} 
    R_{ZZ \, ; WW} \equiv \frac{\mu_{ZZ}}{\mu_{WW}}
                            = 0.951^{+0.531}_{-0.463} \ , 
    \label{R_zz_ww}
\end{equation}
which is rather consistent with the SM. And while $pp \to H \to \tau^+\tau^-$ 
seems to be much smaller than the SM expectations, the current error bars are 
too large for the observable to be of any consequence. 

\section{A brief recapitulation of the pMSSM and its predictions}
\label{sec:pmssm}           
Within the MSSM, the masses and interactions of the Higgs bosons are strongly
affected by loop corrections, in particular from the sector of the
third-generation quarks/squarks. Here, we consider the predictions for
the Higgs bosons in the frame of the phenomenological MSSM (pMSSM) where a non
biased approach to the MSSM parametrisation is adopted and with the
parameters defined and simultaneously varied at the electroweak scale, the
energy scale which the LHC is now probing. 
For the pMSSM fit, the parametrisation is
completely decoupled from the details of the physics responsible for
the breaking of SUSY.  
Only real soft SUSY breaking terms were considered, with all
off-diagonal elements in the sfermion mass terms and trilinear
couplings set to zero, and the first-and second-generation soft terms
equalised\footnote{This set of assumptions are obviously designed 
to conform to the very strict constraints on CP violation as well as 
flavour changing 
neutral current processes, whether in meson mixings or in rare
decays.}, leading to a set of 20 parameters:  
\begin{align}
M_{1,2,3};\;\; m^{3rd \, gen}_{\tilde{f}_{Q,U,D,L,E}},\;\;
m^{1st/2nd \, gen}_{\tilde{f}_{Q,U,D,L,E}}; \;\;A_{t,b,\tau,\mu=e}, \;\;m^2_{H_{u,d}},
\;\;\tan \beta,
\end{align}
where $M_1$, $M_2$ and $M_3$ are the gaugino mass parameters;
$m_{\tilde f}$ are the sfermion mass parameters. $A_{t,b,\tau,\mu=e}$
represent the trilinear scalar couplings while the Higgs-sector
parameters are specified by $m^2_{H_1}$, $m^2_{H_2}$, $\tan \beta$ and
$sign(\mu).$  
The set of observables used for the pre-LHC global
fits~\cite{AbdusSalam:2008uv,AbdusSalam:2009qd} are the 
CDM relic density, electroweak physics observables (EWPOs) and
B-physics observables:
\bea
\underline O &= &\{ m_W,\; \sin^2\, \theta^{lep}_{eff},\; \Gamma_Z,\; \delta
a_{\mu},\; R_l^0,\; A_{fb}^{0,l},\; A^l = A^e,\; R_{b,c}^0,\;
A_{fb}^{b,c},\; A^{b,c}, \\ \nonumber 
& &BR(B \rightarrow X_s \, \gamma),\; BR(B_s \rightarrow \mu^+ \, \mu^-),\; 
\Delta_{0-},\; R_{BR(B_u \rightarrow \tau \nu)},\; R_{\Delta M_{B_s}},\\ \nonumber 
& &\Omega_{CDM}h^2 \}. 
\eea

In the absence of direct data, the pMSSM parameter space is 
weakly constrained in the sense
that only a few of the posterior distributions
(in particular, those that are strongly linked to the Higgs sector) are
approximately prior-independent. The Bayesian fits to data, nonetheless,
have an 
important virtue of indicating 
how effective a particular data is in constraining 
a given model in an unambiguous
manner. For example, if the LHC data rules out a prior-independent pMSSM
prediction, then a large class of MSSM possibilities would have been
ruled out at a go instead of trying many little constructed sub-MSSMs.
This is what we set to explore in our analysis: 
{\it (i)} check if
the Higgs boson discovery data rules out the approximately
prior-independent pMSSM predictions in the Higgs sector and {\it (ii)} 
search for any prior-independent impact of the data on the pre-LHC
posterior samples.

Unlike the case for sparticle spectrum, 
the Higgs sector of the pMSSM is
strongly constrained by the LEP lower limit and the MSSM's
theoretical upper bound on the Higgs mass. This is 
reflected in the results for the pre-LHC Bayesian
fits~\cite{AbdusSalam:2008uv,AbdusSalam:2009qd} where the Higgs 
sector results were found to be
approximately prior-independent. The light CP-even Higgs boson mass
was predicted to be between $117 \gev$ to $129 \gev$ at 95\% Bayesian
confidence level for both logarithmic and flat priors
(i.e., approximately prior independent). This is
consistent with the recent observation of a new state compatible with
the Higgs boson hypothesis near $125 \gev.$ The constraint on the
Higgs sector, in turn, puts a severe constraint on the mass of the scalar
top-quark which plays a crucial role in the radiative contribution to
the Higgs boson mass. As a result, the statistically convergent
Bayesian fits to indirect collider and cold dark matter relic density
data for the pMSSM~\cite{AbdusSalam:2008uv,AbdusSalam:2009qd} reveal
that the scalar top-quarks masses have an approximately prior-independent 
posterior distribution centred around $2 \tev$. 


We assume that the peak discovered by the ATLAS and CMS collaborations
is explained by the pMSSM's light $CP$-even Higgs boson. We then
compute the signal strengths for the production of the Higgs boson 
at\footnote{Although the discovery claims of both ATLAS and CMS are 
based on a combination of $7 \tev$ and $8 \tev$ data, the latter 
has a somewhat weight on account of both higher cross sections as well as
a larger integrated luminosity. Furthermore, the difference between 
the two runs accounts for little in our analysis.}
$8 \tev$ LHC. These are used to check the compatibility and
implications of the discovery data on the pMSSM's both approximately
prior-independent (such as for the scalar top-quarks) and
prior-dependent (as is the case for other sparticle masses) posterior
predictions.  

\subsection{The pre-LHC predictions for $\mu_{X}$ and $R_{ZZ \, ;
    \gamma \gamma}$}  
\begin{figure*}[ht!]
  \begin{center}
    \includegraphics[width=.9\textwidth]{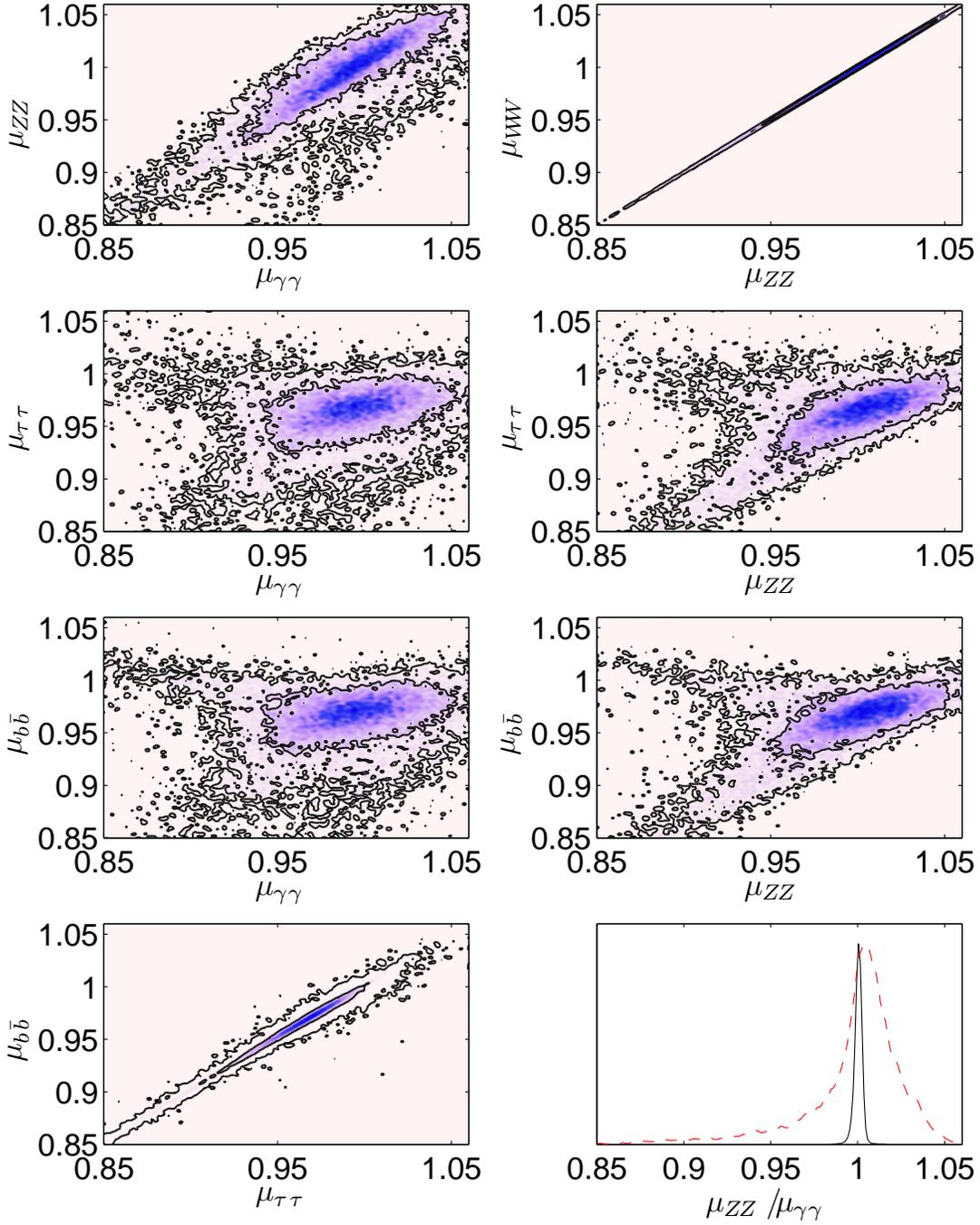}
  \end{center}
  \vspace{-1.0cm}
  \caption{\normalsize 2-D marginalised posterior distributions for the
  log prior pMSSM's Higgs boson signal strengths at the LHC with $8
  \tev$ collision energy. The colour
  spread from light pink to blue represents relative probability spread
  from zero to unity. The inner and outer contours represent 68\% and
  95\% probability regions respectively. The distributions for the
  flat prior sample is approximately the same but with narrower peaks
  around unity. The $\mu_{ZZ} / \mu_{\gamma \, \gamma}$ 1-D
  marginalised posterior distribution are for the log (red/broken
  curve) and flat (black/solid curve) prior pMSSM fits.}   
  \label{fig:discrimL}
\end{figure*}

For each pMSSM model, the Higgs boson production cross section
$\sigma(g \, g \, \rightarrow h)$ and their branching fractions $Br( h
\rightarrow X)$ are computed using the high energy  
physics package \texttt{FeynHiggs}~\cite{feynhiggs}. From these, the
signal strengths $\mu_{X}$ eq.(\ref{muX}) are calculated. 
The pMSSM posteriors for the signal strengths are shown in
Fig.~\ref{fig:discrimL}. The signal
strengths $\mu_{X}$ have a spread encompassing unity, but typically
have somewhat reduced values.  In particular, the signal strengths for
$gg \to h \to b\bar b/ \tau^+ \tau^-$ both, largely, tend to be below
those expected for the standard model. Naively, this might seem
unexpected in view of the fact that the Yukawa coupling in the MSSM is
often larger than in the SM (at least for large $\tan\beta$). However,
this has to be convoluted with the fact that the $gg \to h$ amplitude
is suppressed on account of the cancellation between the top and the 
stop loops. While this cancellation is less effective for super-heavy
stops, such solutions are, typically, less favoured once constraints
such as $\Omega_{CDM}\,h^2$, $BR(B \to X_s \gamma)$ and $\delta a_\mu$
are taken into 
account. Indeed, this could be gleaned from the pre-LHC pMSSM
posteriors as well. The reduction is more pronounced for the log
prior fit's posterior distribution since the sparticle masses are
lighter. Note that $pp \to h \to b\bar b$
would, perhaps, never be seen over the large QCD background, while
both ATLAS and CMS see fewer candidates for $pp \to h \to \tau^+
\tau^-$ than would be expected within the SM. Thus, no conclusion
can be drawn, as yet, from these two modes. 

The $ZZ$ mode strength can be quite consistent with the SM value, with
relatively fewer points 
leading to a suppression. On the other hand, the observed rate too
is marginally smaller than that in SM with an error bar
sufficiently large to preclude any significant
constraint\footnote{It should be noted here that the situation with
  $h \to WW$ is similar, albeit with even larger error bars.}. The
situation with the diphoton rate is more dramatic. While, on the
face of it, even $\mu_{\gamma\gamma}$ can be consistent with unity,
note that this rate largely lies well below the SM one. Part of this
can be attributed to the aforementioned suppression in the
production cross section.  An additional suppression accrues due to
the cancellations between the $W$-loop (dominant contribution within
the SM) and the chargino loops (care must be taken to include the
charged-Higgs loops as well) on one hand and the top- and
stop-loops on the other. More importantly, the observed rate is
significantly above the SM expectation, a situation that is rare within
the MSSM. Thus, if we use $\mu_{\gamma\gamma}$ as a constraint, very
few points would survive in the pMSSM scan, a situation that is
corroborated by Table~\ref{tab:summary}, where we have summarised
the relative number of the model points that survive the various
experimental observations, individually and in combination.

\section{Higgs discovery impact on pMSSM sparticle predictions}
\label{sec:impact}
The Higgs discovery data may be considered to consist of two forms of
observables, namely the Higgs boson mass and the signal strengths. 
Although the pre-LHC data predict $m_h \in [117,129]\ {\rm GeV}$
at 95\% C.L. in correspondence with
large stop masses as every additional increment 
in $m_h$ beyond $m_h = m_Z$ (with only the MSSM particle content)
needs progressively larger stop masses~\footnote{There is an exception
for the ``Natural SUSY'' scenario which allows for light stops and at
the same time satisfying $m_h \sim 125 \gev$. This is possible within
the pMSSM but is beyond our purpose here.}. This immediately implies that 
requiring $m_h = 125-126.5 \gev$ 
would impose significant further constraints on the pre-LHC pMSSM 
posterio distribution, see
Table~\ref{tab:summary}. However, given the about $3 \gev$ errors
from theoretical computations~\cite{mssm_higgs} one could, for
the purpose of the present analysis, consider the Higgs boson mass
range of 
\be
 m_h = 122-128 \gev 
     \label{exptmus_0}
\ee
to be plausible and allowed for the pMSSM predictions.

The experimental values for the Higgs boson signal strengths we use
are shown in Table~\ref{muus}. While the excesses do depend 
on the exact mass bin, it is, nonetheless, useful to combine the 
two data from the two experiments to yield
\be
1.2 \leq \mu_{\gamma \gamma} \leq 2.3, \quad 0.4 \leq \mu_{ZZ} \leq 1.8 \ .
\ee  
While this (along with eq.(\ref{exptmus_0})) does represent a slight 
simplification, this is the best that can be done until the two 
experiments combine their data. 
After applying various constraints including the new discovery
mode signal strengths, the relative amount of 
the model points that survive various combinations of the experimental
cuts eq.(\ref{R_zz_diphot}) and eq.(\ref{exptmus_0}) are summarised in
Table~\ref{tab:summary}.  
\label{sec:gauge}
\begin{table}[!h]
\begin{center}
\begin{tabular}{|c|l|l|l|}
\hline
No. &     Constraint &            Log prior survive & Flat prior
survive \\
\hline
1.&      $m_h=122-128 \gev$&         36.08\%&	     41.24\%\\
2.&      $m_h=125-126.5 \gev$&       9.17\%&	     9.93\%\\     
3.&     $R_{ZZ ; \gamma\gamma}= 0.56 \pm 0.25$ & 0.30\%&  0.00\% (3 points)\\
4.& 2-$\sigma$ $R_{ZZ ; \gamma\gamma}$ & 99.90\%& 100.00\%\\           
5.& 1.67-$\sigma$ $R_{ZZ ; \gamma\gamma}$ & 13.30\%& 0.40\%\\           
6.& $m_h(1)$ \& 1.67-$\sigma$ $R_{ZZ ; \gamma\gamma}$ & 4.67\%& 0.20\%\\           
7.&    $1.2 \leq \mu_{\gamma\gamma} \leq 2.0$ & 0.20\%&	     0.01\% (5 points)\\     
8.&    $0.4 \leq \mu_{ZZ} \leq 1.8 $ &   99.63\%&	     99.96\%\\
9.& $m_h(1)$ \& 1-$\sigma$ $R_{ZZ ; \gamma\gamma}$ & 0.07\%& 0.00\% (0 points)\\           
10.& $m_h(2)$ \& 1-$\sigma$ $R_{ZZ ; \gamma\gamma}$ & 0.00\% (2 points)& 0.00\% (0 points)\\ 
11.& $\mu_{\gamma\gamma}$ \& $\mu_{ZZ}$ & 0.00\% (2 points)& 0.01\% (5 points)\\
\hline 
\end{tabular}
\caption{\normalsize Summary of the relative number of surviving posterior points,
  from the pre-LHC Bayesian global fits of the pMSSM, after imposing
  the Higgs discovery data.} 
\label{tab:summary}
\end{center}
\end{table}

As we have already pointed out, the naive imposition of
$\mu_{\gamma\gamma}$ (or, for that matter, any of the $\mu_{X}$) as a
constraint is fraught with danger in view of the uncertainties in the
production cross section.  On the other hand, the use of $R_{ZZ \, ;
  \gamma \gamma}$ (see eq.(\ref{R_zz_diphot})) largely eliminates
these uncertainties, and given its sizable deviation from unity, is
expected to lead to more robust claims. As Table~\ref{tab:summary}
shows, the latter constraint is not significantly weaker than the
former but, of course, it is far less constraining than the
simultaneous imposition of both $\mu_{\gamma\gamma}$ and $\mu_{ZZ}$.

\subsection{Implications}
The effects of the Higgs data on the pMSSM sparticle posterior
distributions are shown in Fig.~\ref{masses_167sgRNmh} which shows 
the allowed regions of sparticle masses 
for the model points that survive constraint no.~6 in
Table~\ref{tab:summary}. 
Note that the distributions for the surviving points are not meant to
represent the possible outcome of a full-fledged
parameter fit with the Higgs data,
since this would require more statistic and convergence. Observations
about the effect of the different constraints on the sparticle masses
are as follows.
\begin{itemize}
\item {Only \bf $m_h = 122 - 128 \gev$ as a constraint:}
On imposing only this, about 40\% of the posterior
samples survive and the constraint does not affect the shape of the
SUSY-breaking parameters and sparticles posterior distributions. The
only exception is for the strong preference for multi-TeV trilinear
coupling, $|A_t| \sim 5 \tev$. The approximate conservation of the
posterior distributions reflects the consistency of the pre-LHC 
fits with the Higgs boson mass. 

\item {\bf Only $R_{ZZ \, ; \gamma \gamma}$ as a constraint:} 
Although this discriminant shows an experimental central value 
very different from unity, note that it has yet a large error bar. 
Consequently, if we only demand a $2\sigma$ agreement 
with it, then essentially all the points survive for both log
and flat prior samples. This is so because the pMSSM predictions are
centred around unity, which is marginally consistent at $2\sigma$. 
The situation changes dramatically, if we strengthen the demand to a 
$1.67\sigma$ (corresponding to a 90\% C.L. for the 1-D marginalised 
distribution) agreement. Now, 
only about 13.3\% of the log prior and
0.4\% of the flat prior posterior samples survive. Consequently,
we skip the flat
prior samples and continue the analysis with only the log prior. 
The surviving model points show a slight preference
for lighter staus, stau-neutrino and 1st/2nd generation squarks, and
for slightly heavier $\tilde{\chi}_4^0$,
$\tilde{\chi}_2^{\pm}$, stops and $\tilde{b}_2$. All other sparticle
distributions remain approximately unchanged.

\item {\bf $m_h = 122 -128 \gev$ as well as $R_{ZZ \, ; \gamma
\gamma}$ at 90\%C.L.:} 
5.3\% of the log-prior and 0.2\% of the flat-prior 
posterior samples survive. The sparticles' posterior
distributions for the initial log-prior sample set and for the
surviving points are shown in Fig.\ref{masses_167sgRNmh}. As can be
seen from the plots, there is a shift in probability towards
heavier neutralinos (except for the LSP), charginos and stops. On 
the other hand, the data show preference for rather lighter 1st/2nd
generation squarks. The signal strengths' posterior distributions for
the surviving points are shown in Fig.~\ref{fig:LmhDiscMuus}. 

\begin{figure*}
  \begin{center}
    \includegraphics[width=1.\textwidth]{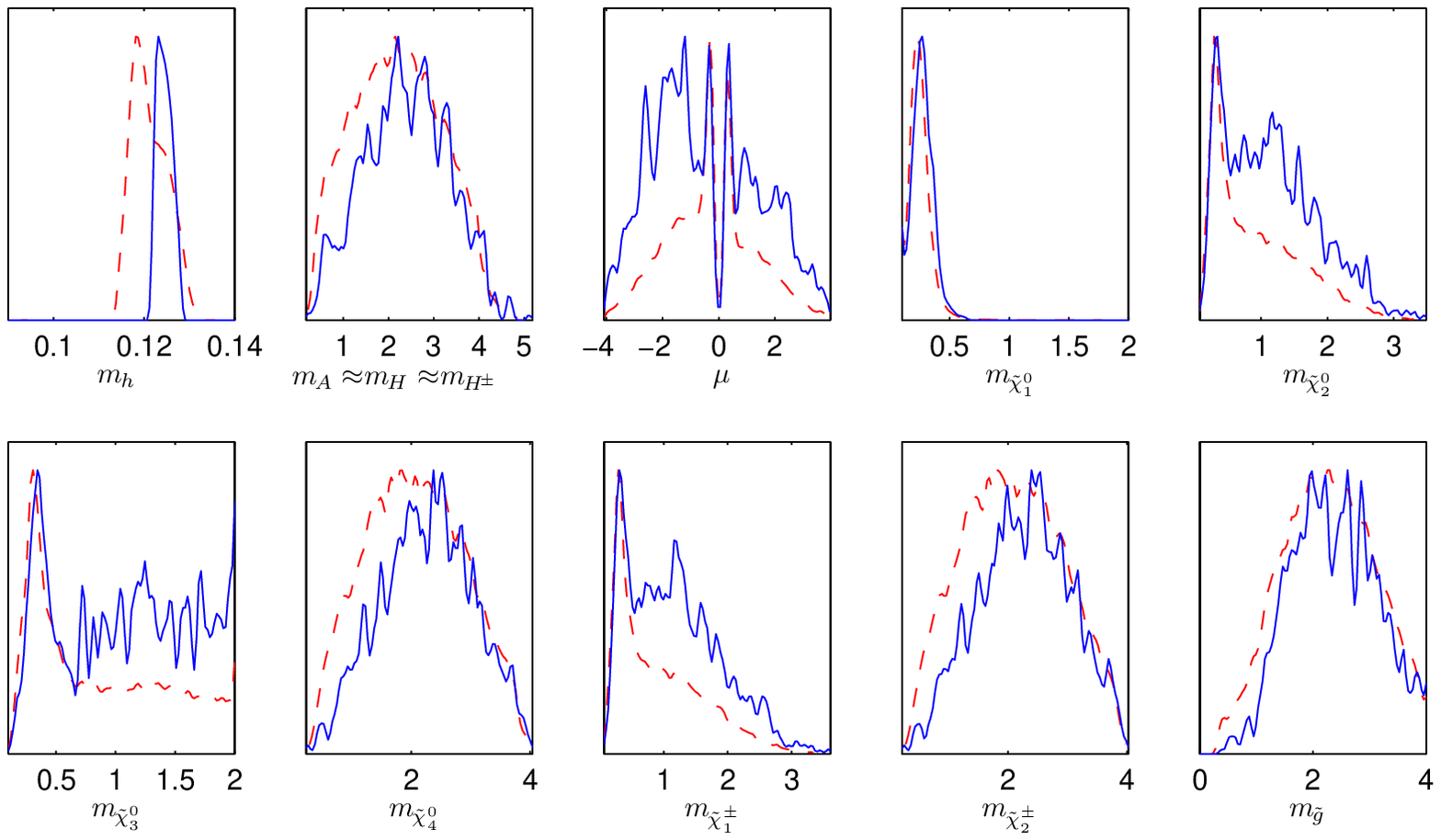}\\
    \includegraphics[width=1.\textwidth]{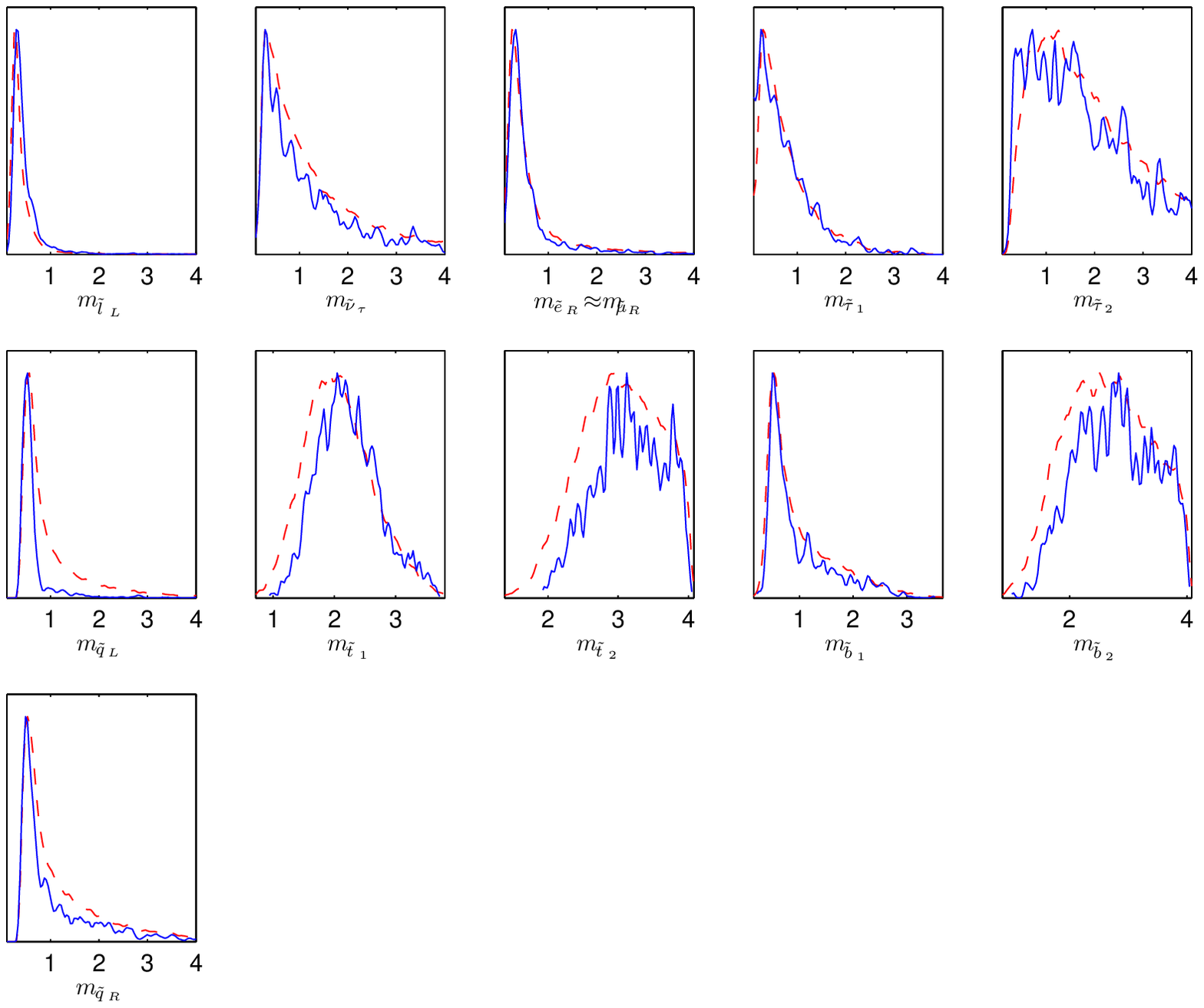}
  \end{center}
  \vspace{-0.6cm}
  \caption{\normalsize The plots compare the log prior pMSSM sparticle
    masses' marginalised 1-dimensional pre-LHC posterior distributions
    (dashed lines) and the surviving parameter regions (solid-blue lines)
    after imposing both $m_h = 122.0 - 128.0 \gev$ and 
    1.67-$\sigma$ $R_{ZZ \, ; \gamma \gamma}$. All the masses are in
    $\tev$ units. The vertical axes represent the relative probability
    weights of the model points.} 
  \label{masses_167sgRNmh}
\end{figure*}

\begin{figure*}
  \begin{center}
    \includegraphics[width=1.\textwidth]{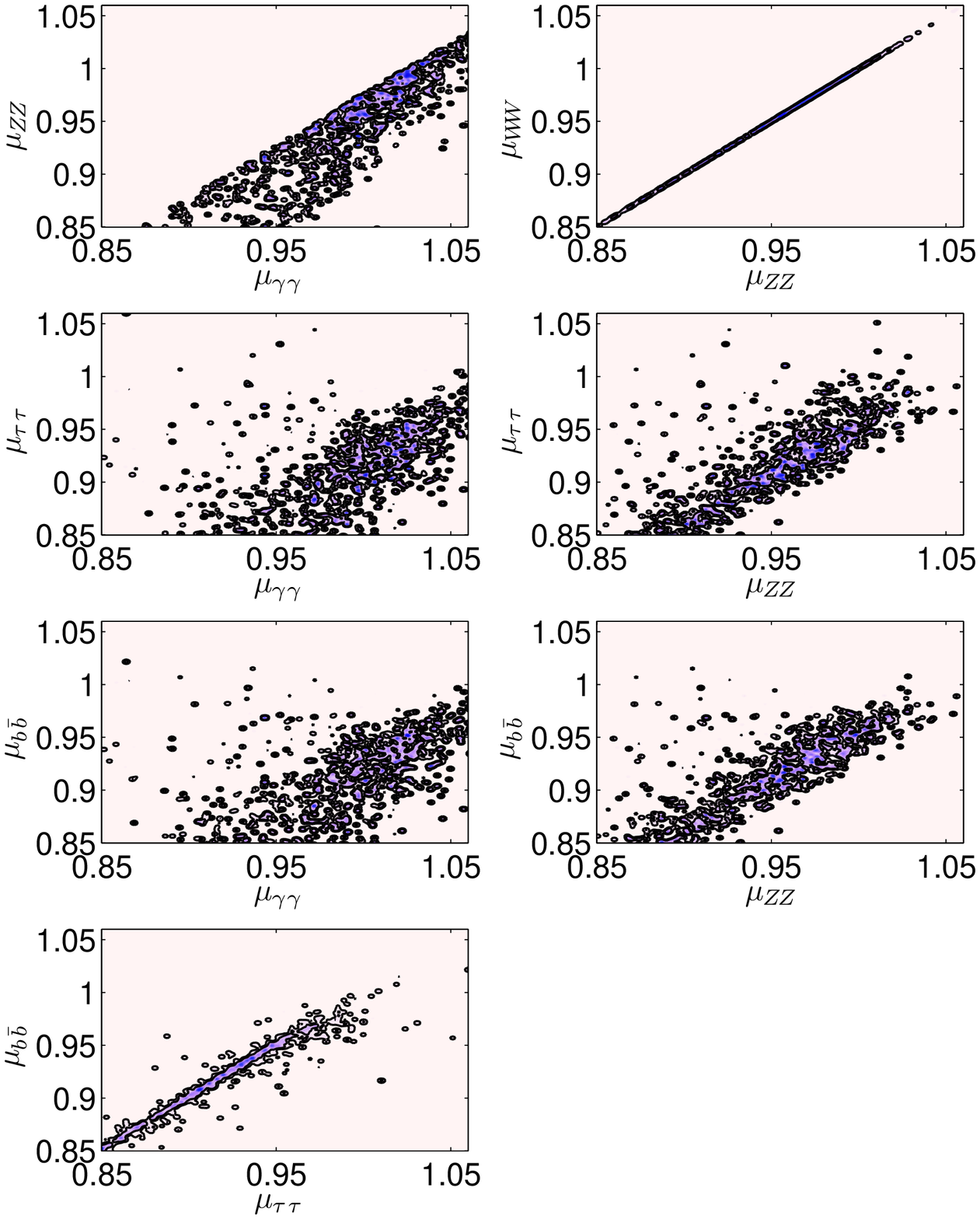}\\
  \end{center}
  \vspace{-0.6cm}
  \caption{\normalsize Marginalised posterior distributions for the
  signal strengths at $8 \tev$  LHC for the pMSSM points that survive
  the $m_h = 122 -128 \gev$ and 1.67-$\sigma$ $R_{ZZ \, ; \gamma
  \gamma}$ constraints. The colour  
  spread from light pink to blue represents relative probability spread
  from zero to unity. The inner and outer contours represent 68\% and
  95\% probability regions respectively. The distributions for the
  flat prior sample is approximately the same but with narrower peaks
  around unity.} 
  \label{fig:LmhDiscMuus}
\end{figure*}

\item {\bf Only $R_{ZZ \, ; \gamma \gamma}$ at $1\sigma$:}
While a demand that a particular observable agree at $1\sigma$, 
statistically speaking, makes little sense, we, nonetheless examine this 
for curiosity. Partly, this exercise is motivated by the realization that, 
the error in,  eq.(\ref{R_zz_diphot}), has most probably been over-estimated, 
and is likely to go down significantly once the common errors have been 
factored out. Furthermore, with more data (already on tapes now), the statistical component would reduce substantially. 
However, working with the currently available data, imposing this implies that
only 0.3\% of the log prior posterior samples survive while all of the
flat prior samples get ruled out. The surviving points show
preference for light (one to few hundred GeVl) staus and stau-neutrinos;
preference for heavier $\tilde{t}_2$, and moderate $\tan \beta
(\lesssim 40)$; and preference for heavy $(\geq 500 \gev)\,
\tilde{\chi}_3^0$ and $(\geq 1000 \gev)\, \tilde{\chi}_4^0$ and
$\tilde{\chi}_2^{\pm}$. The $m_{\tilde t_{1,2}}$ posterior
distribution remain approximately unchanged, mainly peaked around $2
\tev.$ With $m_h = 122 -128 \gev$ and $R_{ZZ;\gamma\gamma}$ at
1-$\sigma$, the few surviving points have the characteristics of:
{\it (i)} having $\tilde{\chi}_1^0 \lesssim m_h/2$, {\it (ii)} having
light ($\lesssim 500 \gev$) quasi-degenerate light 1st/2nd-generation squarks with the
LSP, and {\it (iii)} light stau states. 

\end{itemize}

\section{Summary and conclusions}
The signal strengths for the recently discovered Higgs boson-like
state are used to propose a BSM discriminant, the ratio of diphoton- to
ZZ- channel signal strengths. This has the advantage that 
sources of theoretical uncertainties cancel almost completely unlike
the case for the individual signal strengths. The discriminant is
applied  to the 
posterior distributions for a pre-LHC Bayesian fit to the pMSSM,
sorting for the impact of the discovery data on the sparticle mass
predictions. The data provides mainly two classes of information; one
about the mass and the other about the signal strength of the new
state.

The pre-LHC pMSSM fits predict an approximately prior-independent
Higgs boson mass to be between $117 \gev$ and $129 \gev$ at 95\%
Bayesian confidence level and, hence, are in agreement with the recent
discovery of a new state around $126 \gev.$ The pMSSM's signal
strengths are narrowly distributed around unity, as shown in
Fig.~\ref{fig:discrimL} suggesting compatibility with the SM
expectations. These are in agreement with the experimentally
determined values, to within 2$\sigma$ (or 95\% C.L.) for the diphoton
channel.  This also holds for the aforementioned discriminant, simply
as a consequence of the as yet large error bars associated with it. In
other words, at this level, the effect of the signal strengths is not
yet conclusive, and better precisions in the results is necessary for
making distinctive conclusions.

On the other hand, if we demand that the MSSM expectation of the
discriminant agree with the data to within $1\sigma$, almost none,
viz. 0.0\% (3 model points) of the flat prior and some 0.3\% (125
model points) of the log prior posterior samples survive. Indeed, even
at 90\% C.L., the constraints are quite strong.  The surviving regions
of the parameter space typically are associated with lighter stau
states and lighter 1st/2nd-generation squarks, whereas the stops, the
charginos and the (non-LSP) neutralinos tend to be heavy.  The
prediction for light squarks is apparently in tension with limits from
SUSY searches. However, it should be realized that the purpose here is
to look for the directions in sparticle mass space to which the Higgs
data pull, and the tension is not an overwhelming one.  As for the
stau states, the current direct LHC limits from either (Drell-Yan
like) pair-production or staus from weak gaugino decays are weak.
Future limits on the stau mass would be important constraints for the
1$\sigma$ $R_{ZZ \, ; \gamma \gamma}$ surviving points.

In conclusion, assuming an R-parity conserving MSSM, the results of
our analyses imply that 
if the diphoton signal strength anomaly persists even 
after the accumulation of more
LHC data, then it is very likely that there exist 
light sparticles which have to-date escaped detection. This could be
due to the low magnitude of missing energy in the case of light
(compared to the experimental missing transverse energy cuts)
neutralinos and/or due to the
quasi-degenerate nature of the spectrum which 
leads to soft 
jets~\cite{AbdusSalam:2011hd,AbdusSalam:2011fc} that would be buried
under QCD background~\footnote{It is not obvious if the compressed
SUSY scenario~\cite{LeCompte:2011cn} considered within a
modified-CMSSM/mSUGRA benchmarking models can survive with Higgs
boson mass around $126 \gev$~\cite{:MET}.}. Given the relatively low
statistics for the surviving points, it 
is now time for updating statistically robust trends for MSSM physics
by performing new pMSSM explorations and global fits to the
post-discovery LHC data. 
Before this, short-range outlook indicates that performing a
similar study, but for the impact of the LHC SUSY limits on the pMSSM
posterior distributions is well within reach~\cite{:2012wip}. The same
is the case for checking the impact on the freedom of choice for
considering the LSP as being either the only or a partial
source for CDM relic density within the pMSSM
framework~\cite{AbdusSalam:2010qp}. 
Other interesting directions include searching for plausible
effects of non-diagonal flavour on the pMSSM predictions and for
computing the impacts of the discovery data and SUSY limits on a
phenomenological next-to-MSSM, the pNMSSM.

\section*{Acknowledgments}
We thank F.~Quevedo for useful comments and
discussions. DC thanks the Abdus Salam ICTP for hospitality during the 
period when this project was conceptualised and started.

\end{document}